\documentclass[symmetry,article,accept,pdftex,moreauthors]{Definitions/mdpi} 

\usepackage{graphicx}
\usepackage{amssymb}
\usepackage{amsmath}
\usepackage{tensor}
\usepackage{nicefrac}
\usepackage{amsthm}
\usepackage{mathrsfs}

\usepackage{color,xcolor}
\firstpage{1} 
\pubvolume{16}
\issuenum{7}
\articlenumber{881}
\pubyear{2024}
\copyrightyear{2024}
\externaleditor{\textls[-25]{Academic Editor: Kazuharu Bamba}}
\datereceived{7 June 2024} 
\daterevised{2 July 2024} 
\dateaccepted{7 July 2024} 
\datepublished{11 July 2024} 
\hreflink{\textls[-15]{https://doi.org/10.3390/sym16070881}} 

\Title{Extension of Buchdahl's Theorem on Reciprocal Solutions}

\TitleCitation{\textls[15]{Extension of Buchdahl's Theorem on Reciprocal}\linebreak Solutions}

\newcommand{\orcidJPMimoso}{{\href{https://orcid.org/0000-0002-9758-3366}{\orcidicon}}} 
\newcommand\orcidFrancisco{{\href{https://orcid.org/0000-0002-9388-8373}{\orcidicon}}}

\Author{David S. Pereira 
 $^{1}$, José Pedro Mimoso $^{1,2,}$*\orcidJPMimoso \, and Francisco S. N. Lobo $^{1,2,}$*\orcidFrancisco}

\AuthorNames{David S. Pereira, José Pedro Mimoso, Francisco S. N. Lobo}

\AuthorCitation{Pereira, D.S.; Mimoso, J.P.; Lobo, F.S.N.}

\address{%
$^{1}$ \quad Instituto de Astrof\'{\i}sica e Ci\^{e}ncias do Espa\c{c}o, Faculdade de
Ci\^encias da Universidade de Lisboa, \linebreak Campo Grande, Edif\'{\i}cio C8,
P-1749-016 Lisbon, Portugal; davidsp@alunos.fc.ul.pt\\
$^{2}$ \quad Departamento de F\'{i}sica, Faculdade de Ci\^{e}ncias da Universidade de Lisboa, Campo Grande, Edif\'{\i}cio C8, P-1749-016 Lisbon, Portugal}

\corres{Correspondence: jpmimoso@fc.ul.pt (J.P.M.); fslobo@fc.ul.pt (F.S.N.L.)}

\abstract{Since the development of Brans--Dicke gravity, it has become well-known that a conformal transformation of the metric can reformulate this theory, transferring the coupling of the scalar field from the Ricci scalar to the matter sector. Specifically, in this new frame, known as the Einstein frame, Brans--Dicke gravity is reformulated as General Relativity supplemented by an additional scalar field. In 1959, Hans Adolf Buchdahl utilized an elegant technique to derive a set of solutions for the vacuum field equations within this gravitational framework. In this paper, we extend Buchdahl's method to incorporate the cosmological constant and to the scalar-tensor cases beyond the Brans--Dicke archetypal theory, thereby, with a conformal transformation of the metric, obtaining solutions for a version of Brans--Dicke theory that includes a quadratic potential.  More specifically, we obtain synchronous solutions in the following contexts: in scalar-tensor gravity with massless scalar fields,  Brans--Dicke theory with a quadratic potential, where we obtain specific synchronous metrics to the Schwarzschild--de Sitter metric,  the Nariai solution, and a hyperbolically foliated solution.}


\begin{document}



\section{Introduction}


The Brans--Dicke theory \cite{Brans:1961sx,Dicke:1961gz}, proposed by Carl Brans and Robert H. Dicke in 1961, is an alternative to Einstein's General Theory of Relativity (GR). It introduces a varying gravitational ``constant'' by incorporating a scalar field $\phi$ alongside the tensor field $g_{\mu\nu}$ of GR. 
The strength of the coupling between the scalar field and the metric tensor is dictated by the Brans--Dicke parameter $\omega$. As $\omega \rightarrow \infty$, Brans--Dicke theory approaches GR, with the scalar field becoming less influential.
The theory can be tested against observations, such as the perihelion shift of Mercury, light deflection by the Sun, and time delay of radar signals \cite{Will:2014kxa}. Constraints on $\omega$ from these tests have required $\omega$ to be very large, making the theory increasingly similar to GR in the observed regime. 
However, the Brans--Dicke framework can be generalized to Scalar-Tensor theory by allowing for more general coupling functions and potential terms for the scalar field, thereby accommodating a broader range of theoretical possibilities and observational phenomena (we refer the reader to \cite{Fujii:2003pa} for more details). This generalization enables a more flexible description of gravitational interactions and has important implications for cosmology, particularly in explaining the accelerated expansion of the universe and testing alternative models of gravity (see \cite{Faraoni:2004pi,Clifton:2011jh} for details).

An interesting application to Scalar-Tensor (ST) theory is Buchdahl's work on reciprocal static metrics and scalar fields within GR, which offered a profound exploration of the mathematical structures underlying gravitational fields. In 1954, Buchdahl initiated the study of these metrics, which exhibit a symmetry under a reciprocal transformation of the radial coordinate \cite{Buchdahl:1954}. He extended this analysis, in 1956, by presenting explicit solutions to the gravitational field equations that remain invariant under these transformations \cite{Buchdahl:1956zz}. Later, in 1959, Buchdahl further investigated the role of scalar fields in this context \cite{Buchdahl:1959nk}, demonstrating how these fields interact with and modify the reciprocal static metrics.  In fact, his powerful technique enabled him to construct the most general two-parameter family of solutions of the static field equations, which was later generalized to $D$-dimensions~\cite{Xanthopoulos:1989kb}. In fact,  Buchdahl's procedure produced the explicit expressions for the solutions sought by Bergmann and Leipnik \cite{Bergmann:1957zza}, who discovered a two-parameter family of solutions, of which the Schwarzschild solutions form a one-parameter subfamily. Here, the metric tensor and scalar field were only implicitly expressed as algebraic and logarithmic functions of the radial coordinate $r$,  except for special one-parameter subfamilies where explicit expressions were obtainable.
Indeed, this work collectively highlight Buchdahl's significant contributions to understanding the complex solutions of Einstein's equations, revealing the subtle relationships between scalar fields and spacetime geometry.

Buchdahl's work has inspired much research over the years \cite{Rao:1972kw,Bekenstein:1974sf,Cadoni:1991ne,Wiltshire:1992jc,Mignemi:1988qc,Schmoltzi:1991fb,Jetzer:1992np,Poletti:1994ff,Fonarev:1994xq,Park:1996zf,Bhadra:2001fx,Schunck:2003kk,Bhadra:2005mc,Capozziello:2024ucm,Capuano:2023yyh,Kobayashi:2014eva,Charmousis:2015aya,Nguyen:2023qux,Nguyen:2023fzk,Nguyen:2024efh} (we refer the reader to   \cite{Faraoni:2021nhi} for a recent review).
Here, we extend Buchdahl's method to incorporate the cosmological constant, thereby, with a conformal transformation of the metric, obtaining solutions for a version of Brans--Dicke theory that includes a quadratic potential. We also devise an application of the Buchdahl reciprocity mechanism to derive solutions of Brans--Dicke theory with  a stiff matter content, conveyed by a massless scalar field. This constitutes a novel realization of the Buchdahl's correspondence between solutions overcoming the restriction to vacuum settings. 

This paper is organized in the following manner: In Section \ref{Symmetries of nested conformal transformations}, we introduce the fundamental concepts of conformal transformations, synchronous metrics, and Buchdahl's reciprocity procedures.  In Section \ref{Solutions from Buchadahl's method with Lambda}, we extend Buchdahl's technique in the presence of a cosmological constant, and find a one-parameter set of static solutions.  In Section \ref{Sec4:applications}, we consider specific applications, namely, to 
scalar-tensor gravity with massless scalar fields, to Brans--Dicke theory with a quadratic potential, where we obtain specific synchronous solutions to the Schwarzschild--de Sitter metric,  the Nariai solution, and a hyperbolically foliated solution.

\section{Symmetries of Nested Conformal Transformations }\label{Symmetries of nested conformal transformations}

\subsection{Conformal Transformation}

Consider a given metric $\bar g_{ab}$  supposedly a solution of Einstein's field equations. Consider that it transforms under the following rescaling:  
\begin{equation}
\bar g_{ab} \longrightarrow {g}_{ab} = \Omega^2 \; \bar g_{ab}\,,
 \label{eq:ct}
\end{equation}
where $\Omega$ is a regular (analytic) function of the space-time coordinates. 
We are introducing a scaling which may vary from point to point, and as  a  
consequence,  the relevant geometric quantities involved in the 
gravitational theory are modified.  The connections become
\begin{equation}
{{\Gamma^a}}_{bc} = {\bar \Gamma^a}_{bc} + \delta^a_b  
\partial_c (\log{\Omega}) +\delta^a_c \partial(\log{\Omega}) -g_{bc} 
\partial^a (\log{\Omega}) \,.
 \label{eq:cconect}
\end{equation}

It is immediately apparent that the scalar function involved in the 
definition of the conformal transformation induces the presence of an 
additional term in the transformed connection. We write
\begin{equation}\label{Connections relation}
{{\Gamma^a}}_{bc}={\bar \Gamma^a}_{bc} + {\gamma^a}_{bc} \,,
\end{equation}
and this carries to the Ricci tensor in the following manner:
\begin{equation}
{{R}}_{ab} = {\bar{R}}_{ab}  +  P_{ab} \,,
\end{equation}
where the $P_{ab}$ tensor is the ``Ricci'' tensor constructed from the 
connection  ${\gamma^a}_{bc}$. The form of this extra $D$-dimensional Ricci tensor is  
\cite{Wald:1984rg}
\begin{eqnarray}
P_{ab} & = & -  (D-2) \nabla_a \nabla_b \log{\Omega } -  g_{ab} 	\nabla^c \nabla_c \log{\Omega} 
		\nonumber \\ 
 	&  & + (D - 2) \nabla _a \log{\Omega}  \nabla_b \log{\Omega} 
 		 \\
	& & - (D-2) \nabla^c \log{\Omega} \nabla_c  \log{\Omega} \,, \nonumber
\end{eqnarray}
where the trace $P$ yields
\begin{eqnarray}\label{Trace of Ricci transformed}
P   \equiv   {P^a}_a  & = &  \Omega^{-2}\; \bar R  - \Omega^{-2}  \Big[2 (D-2) \nabla^a \nabla_a \log{\Omega} 
	 \nonumber \\
     &&  -(D-2) (D-1) 
	\nabla^a  \log{\Omega} \nabla_a  \log{\Omega} \Big]\,.
\end{eqnarray}

From the conformal transformation (\ref{eq:ct}),  there is no prescription 
for transforming  objects such as scalar fields, vectors, or general 
tensors. However, it seems appropriate to assume that these objects will
undergo scale transformations similar to (\ref{eq:ct}), such as
\begin{equation}
\psi  \longrightarrow  \bar{\psi} = \Omega^s\;  \psi \,,
\end{equation}
or 
\begin{equation}
T_{ab} \longrightarrow  {\bar{T}}_{ab} = \Omega^p \; T_{ab} \,.
\end{equation}

Using these elements, we obtain for the transformed Einstein field equations
the following expression:
\begin{eqnarray}
{{R}}_{ab} - \frac{1}{2}\, {{g}}_{ab}\, R & = & {\bar{R}}_{ab} - 
\frac{1}{2}\, g_{ab}\, \bar R -  (D-2)\left[  \nabla_a \nabla_b \log{\Omega } - g_{ab}\;
 \nabla_c \nabla^c  \log{\Omega }\right]  \nonumber \\
& & - \nabla _a \log{\Omega}  \nabla_b \log{\Omega}
\frac{3-D}{2}\, g_{ab}\; \nabla^c \log{\Omega} \nabla_c  \log{\Omega}\,.
\end{eqnarray}
where $R_{ab}-\frac{1}{2}\,g_{ab}\,R= 8\pi G\, T_{ab}$.

\subsection{Synchronous Metric}\label{synchronous metric}

Consider the  synchronous metric, with $D=4$:
\begin{equation}
{\rm d}s^2=-{\rm d}t^2 +\gamma_{ij} {\rm d}x^i {\rm d}x^j\,,
\label{sync_metric}
\end{equation}
with $i,j,=1,2,3$.
The non-trivial components of the connection are given by
%
\begin{eqnarray}
{\Gamma^0}_{ij} &=& \frac{1}{2}\,\partial_0 \gamma_{ij}  \; \equiv 2K_{ij} \nonumber \\ 
{\Gamma^i}_{0j} &=& \frac{1}{2}\,\gamma^{ik}\,\partial_0 \gamma_{kj}  = 2{K^i}_{j} \\
{\Gamma^i}_{jk} &=& \frac{1}{2}\,\gamma^{im}\, \left(\partial_j \gamma_{mk} + \partial_k \gamma_{jm}- \partial_m \gamma_{jk} \right)  \equiv {\Lambda^i}_{jk} \; , \nonumber
\end{eqnarray}
where $K_{ij} $ is the extrinsic curvature. From these we derive the Ricci tensor: 
\begin{eqnarray}
{R^0}_0  &=&  - \frac{1}{2}\,\frac{\partial }{\partial t} {K^j}_j  - \frac{1}{4}\, {K^j}_m {K^m}_j \\
{R^0}_j &=& \frac{1}{2}\, \left( \nabla_m {K^m}_j  -  \nabla_j {K^m}_m \right)\\
{R^i}_j  &=&  {^\ast {\hspace{-2pt}}R^i}_j  - \frac{1}{2\sqrt{\gamma}}\, \frac{\partial }{\partial t}\left(\sqrt{\gamma} {K^i}_{j} \right)
\end{eqnarray}
and ${^\ast {\hspace{-2pt}}R^i}_j$ is the 3-dimensional Ricci tensor associated with the metric induced on the spatial hypersurfaces $\gamma_{jk}$. $\gamma$ is its determinant.

Thus, if the metric is static, an isometry exists, characterized by a time-like Killing vector orthogonal to the spatial hypersurfaces. Then  $\gamma_{jk}$  does not depend on the time coordinate $t$,  so that, on the one hand, $K_{ij} =0$, and on the other, we have $ {R^0}_0=0$, and ${R^i}_j  =  {^\ast {\hspace{-2pt}}R^i}_j $. This implies that $R={^\ast {\hspace{-2pt}}R}$.
It follows that, if both  ${R^a}_b=0$ and $R=0$, so are ${^\ast {\hspace{-2pt}}R^i}_j=0=R={^\ast {\hspace{-2pt}}R}$. In other  words, if the metric $\bar g_{ab}$ describes a vacuum solution, so does~${\gamma_{ij}}$.

\subsection{Buchdahl's Reciprocity}

The idea underlying  Buchdahl's reciprocity can then be perceived in the following way. If we consider a metric $g_{ab}$ that has the form 
\begin{equation}
g_{ab} = {\rm diag}\,\left( e^{\nu(r)},h_{ab}(r,\theta,\psi)\,\right)\,,
\end{equation}
the line element can be cast as
\begin{eqnarray}
\textrm{d}s^2 &=& e^{\nu(r)}\,  \left( - {\rm d}t^2 + e^{-\nu(r)}\, h_{ab}(r,\theta,\psi)\,  \textrm{d}x^a \textrm{d}x^b \right) 
	\nonumber \\
 & = &  e^{\nu(r)}\,  \left(-{\rm d}t^2 +\gamma_{ij} \,{\rm d}x^i {\rm d}x^j\right) \; ,
\end{eqnarray}
where $\gamma_{ij} = e^{-\nu(r)}\, h_{ab}(r,\theta,\psi)$ is independent of the time coordinate. 

Therefore,  if the synchronous metric were to be a  solution of Einstein's vacuum field equations, and $g_{ab}$ is the conformal rescaling of the latter synchronous metric, with 
\begin{equation}
\Omega^2 (r) =e^{\nu(r)}\; ,
\end{equation} 
then from Equation (\ref{eq:ct}), we verify that $g_{ab}$ is the solution of the vacuum, plus a scalar field, which coincides with the re-scaling factor $\Omega$.

One question that is interesting to address is how this result can be understood in terms of symmetries. When the synchronous metric (\ref{sync_metric}) describes a static vacuum solution, then it admits, as mentioned above, a time-like Killing vector that is orthogonal to the hypersurfaces endowed with the metric $\gamma_{ij}$. The Killing vector $X^a$ satisfies the equation
\begin{equation}
\bar \nabla_a X_b+\bar \nabla_b X_a=0\; ,
\end{equation}
where $\bar \nabla_a = h^b_a\,\nabla_b$ is the projected $\nabla$ operator onto the spatial part of $g_{ab}$.  The conformal transformation preserves angles;  hence, if the  Killing field is orthogonal to $\gamma_{ij}$, it will also be orthogonal to $h_{ab}$. 

Thus, the objective here is to analyze how the Killing vector field is transformed from one conformally related metric to the other, in order to assess whether a scalar field that is a Killing of one of the metrics is also a Killing of its image under the conformal transformation. Furthermore, we will explore the cosmological constant in this context, and extend Buchdahl's result to the scalar-tensor cases beyond the Brans--Dicke archetypal~theory.



\section{Solutions from Buchdahl's Method with $\Lambda$}\label{Solutions from Buchadahl's method with Lambda}

To obtain solutions to the field equations, consider the following Lagrangian given by
\begin{equation}\label{Lagrangian Lambda}
\mathcal{L} =  R - 2\Lambda - \mu g^{ab} \partial_a\psi \partial_b \psi \, ,
\end{equation}
where $R$ is the Ricci scalar of the 4-dimensional metric $g_{ab}$, $\mu$ is a constant parameter, and $\psi$ is a scalar field. In the following, we will deduce in detail Buchadahl's method carried out in \cite{Buchdahl:1959nk}, with the inclusion of the cosmological constant $\Lambda$.

\subsection{Buchdahl's Method}

To start, consider a metric $g_{ab}$, with signature $(-,+,+,+)$, that has a cyclic coordinate $x^s$ such that 
\begin{equation}
    \partial_s g_{ab}= 0\,, \qquad
    g_{ks}=0 \, ,
\end{equation}
where  $k,l=(0,1,2,...,s-1,s+1,...,D\geq 3)$. With this defined, consider now the following line element:
\begin{equation}
    \textrm{d}s^2 = g_{ss} (\textrm{d}x^s)^2 + \hat{g}_{kl}\textrm{d}x^k \textrm{d}x^l \, ,
\end{equation}
which can be written in the following compact notation:
\begin{equation}\label{métrica estatica}
	g_{ab} \equiv ( g_{ss},\hat{g}_{kl}) \ ,
\end{equation} 
where $g_{ss}$ is associated with the cyclic coordinate ($x^s$), which in our case is considered to be the time coordinate $x^0$;  $\hat{g}_{kl}$ is a 3-dimensional metric, of the Riemann space $\hat{\mathcal{M}}_{3}$, independent of the $x^s$-coordinate.

Without loss of generality, one can consider $\hat{g}_{kl} = e^{2\sigma}g_{kl}$. More specifically, $\hat{g}_{kl}$ is conformal to the $g_{kl}$ metric, with the Riemann space $\hat{\mathcal{M}}_{3}$ being conformal to the Riemann space $\mathcal{M}_{3}$, where $\sigma$ is a scalar function that is independent of 
 the cyclic coordinate. In the same line of thought, one can define $g_{00}=-e^{2\gamma}$ where $\gamma$ is also a scalar function independent of the cyclic coordinate. In this way,  $g_{ab}$ can be written as 
\begin{equation}\label{metric static g_ab}
    g_{ab} = \left(-e^{2\gamma},e^{2\sigma}g_{kl} \right) = e^{2\gamma}\left(-1,e^{2\omega} g_{kl}\right) = e^{2\gamma} \Bar{g}_{ab} \, ,
\end{equation}
where
\begin{equation}\label{metric static g_ab 2}
      \Bar{g}_{ab} = \left(-1,e^{2\omega} g_{kl}\right) = (-1,\Bar{g}_{kl}) \, ,
\end{equation}
with $\omega = \sigma - \gamma$. The $\Bar{g}_{kl}$ metric is the conformal metric of ${g}_{kl}$ associated with $e^{2\omega}$, whose space is the Riemann space $\Bar{\mathcal{M}}_{3}$, which is conformal to the Riemann space $\mathcal{M}_{3}$. 

As verified in Section \ref{Symmetries of nested conformal transformations}, in the presence of a rescaling of the metric, there is a relationship between the Ricci scalar of the original metric, $g_{ab}$ ($ R [ g_{ab} ]$), and the Ricci scalar of the transformed metric, $\Bar{g}_{ab}$ ($R[\Bar{g}_{ab}]$), as seen in Equation \eqref{Trace of Ricci transformed}.  This transformation leads to the following relation:
\begin{equation}\label{Ricci metric g_ab 1}
    R [ g_{ab} ] = e^{-2\gamma}\left\{ \Bar{R}[\Bar{g}_{ab}] + 6 \Bar{\square} \gamma + 6\Bar{g}^{ab}\partial_a\gamma \partial_b\gamma \right\} \, ,
\end{equation}
where $\Bar{\square}$ is the d'Alembert operator defined as $\Bar{\square} \equiv \Bar{g}^{ab}\Bar{\nabla}_a \Bar{\nabla}_b$. Due to $\gamma$ being independent of the time coordinate, and taking into account Equation \eqref{metric static g_ab 2}, the covariant derivatives present in the d'Alembert operator $\Bar{g}^{ab} \Bar{\nabla}_a \Bar{\nabla}_b$ reduces to $\Bar{g}^{kl} \bar{\nabla}_k \Bar{\nabla}_l$, where $\bar{\nabla}$ are the covariant derivatives with respect to the metric $\Bar{g}_{ab}$. Thus, Equation \eqref{Ricci metric g_ab 1} can be written as
\begin{equation}\label{Ricci metric g_ab 2}
    R [ g_{ab} ] = e^{-2\gamma}\left\{ \Bar{R}[\Bar{g}_{ab}] +  6\Bar{g}^{kl} \Bar{\nabla}_l\Bar{\nabla}_k\gamma + 6\Bar{g}^{kl}\partial_k \gamma \partial_l \gamma \right\} \, .
\end{equation}

Due to the synchronous nature of the metric \eqref{metric static g_ab 2}, which can be rewritten as
\begin{equation}
    \textrm{d}s^2 = -\textrm{d}t^2 + \Bar{g}_{kl}\, \textrm{d}x^k \textrm{d}x^l \, ,
\end{equation}
if $\Bar{g}_{kl}$ does not depend on the isolated coordinate, $x^s$, then the Ricci scalar associated with the metric $\Bar{g}_{ab}$ is equal to the one associated with $\Bar{g}_{kl}$, as seen in the previous section. With this equality, one can write the Ricci scalar associated with $g_{ab}$ in terms of the Ricci scalar associated with $g_{kl}$. 
From Equation \eqref{Trace of Ricci transformed}, this relation is given by
\begin{equation}\label{Ricci g_kl}
    \Bar{R}[\Bar{g}_{ab}]=\Bar{R}[ \Bar{g}_{kl} ] = e^{-2\omega}\left\{ {R}[{g}_{kl}] + 2 g^{kl} \nabla_l\nabla_k\omega + 2 {g}^{kl}\partial_k \omega \partial_l \omega \right\} \, ,
\end{equation}
allowing one to rewrite Equation \eqref{Ricci metric g_ab 2} as
\begin{equation}\label{Ricci metric g_ab 4}
    R [ g_{ab} ] =\ e^{-2\gamma-2\omega} \Big\{{R}[{g}_{kl}] + 2  {g}^{kl} \nabla_l\nabla_k\omega + 2{g}^{kl}\partial_k \omega \partial_l \omega + 6  {g}^{kl} \nabla_l\nabla_k\gamma + 6 {g}^{kl}\partial_k \gamma \partial_l \gamma \Big\} \, .
\end{equation}

To simplify this relation, it is essential to recognize that there are two distinct covariant derivatives at play with respect to $\gamma$: one associated with the metric $\Bar{g}_{kl}$, $\Bar{\nabla}_l\Bar{\nabla}_k\gamma$, and another related to the metric ${g}_{kl}$, $\nabla_l\nabla_k \omega$ (where $\omega = \sigma - \gamma$). Therefore, in order to simplify the expression for the Ricci tensor, we must express $\Bar{\nabla}_l\Bar{\nabla}_k$ in terms of $\nabla_l\nabla_k$. Given that $\Bar{\nabla}_k\gamma = \nabla_k\gamma = \partial_k \gamma$, and considering the conformal relationship between the metrics as described in Equation \eqref{Connections relation}, one can write
\begin{eqnarray}
    \Bar{\nabla}_l\Bar{\nabla}_k\gamma &=&  \partial_l\partial_k\gamma - \Bar{\Gamma}^i_{kl} \partial_i\gamma 
    \nonumber \\
     &= & \partial_l\partial_k\gamma - {\Gamma}^i_{kl} \partial_i\gamma - \delta_k^i\partial_l \omega \partial_i\gamma - \delta_l^i\partial_k \omega\partial_i\gamma + g_{kl} g^{im} \partial_m\omega \partial_i \gamma
        \\
    &=& \nabla_l \nabla_k \gamma - \delta_l^i\partial_k \omega \partial_i \gamma + g_{kl} g^{im} \partial_m\omega \partial_i\gamma \, . \nonumber
\end{eqnarray}
Reintroducing this new expression in Equation \eqref{Ricci metric g_ab 4} and using the definition of $\omega$, the final result is given by 
\begin{equation}\label{Ricci metric g_ab final}
    R [ g_{ab} ] = \ e^{-2\sigma} \left\{{R}[{g}_{kl}] + 2{g}^{kl}\left( 2\nabla_l \nabla_k \sigma + \nabla_l \nabla_k \gamma + \partial_k \sigma\partial_l \gamma + \partial_k \sigma\partial_l \sigma+\partial_k \gamma\partial_l \gamma \right) \right\} \, .
\end{equation}

Working with conformal transformations, the determinant of the metric $g_{ab}$ can be expressed as a conformal factor times the determinant of the $g_{kl}$ metric. Thus, $\sqrt{-|g_{ab}|}$, a tensor density of weight 1, can be related to $\sqrt{-|\Bar{g}_{ab}|}$ in the following way:
\begin{equation}\label{determimante metrica}
    \sqrt{-|g_{ab}|} = (e^{\gamma})^D \sqrt{ -|\Bar{g}_{ab}|} = e^{4\gamma}\sqrt{ |\Bar{g}_{ab}|}\, .
\end{equation}

Due to the way that the metric $\Bar{g}_{ab}$ is constructed, one has $\sqrt{- |\Bar{g}_{ab}|} = \sqrt{-|\Bar{g}_{kl}|}$, so that, due to conformal relation between the metric $\Bar{g}_{kl}$ and ${g}_{kl}$, this 
  can be expressed as 
\begin{equation}
    \sqrt{-|\Bar{g}_{ab}|} = \sqrt{-|\Bar{g}_{kl}|} = e^{3\omega}\sqrt{|{g}_{kl}|} \, ,
\end{equation}
allowing one to rewrite Equation \eqref{determimante metrica} as
\begin{equation}\label{relação determinantes g}
    \sqrt{-|g_{ab}|} = e^{3\sigma+\gamma} \sqrt{-|{g}_{kl}|} \, .
\end{equation}

Multiplying Equation \eqref{Ricci metric g_ab final} by $\sqrt{-|g_{ab}|}$, one obtains that the $g_{ab}$ metric has a scalar curvature density $\mathcal{R}$ ($R\sqrt{-|g_{ab}|}$) given by
\begin{equation}
    \mathcal{R} = e^{\gamma + \sigma} \sqrt{-|{g}_{kl}|}\left\{ R[{g}_{kl}] + 2g^{kl}\left(2\nabla_l \nabla_k \sigma+\nabla_l \nabla_k \gamma+ \partial_k\sigma\partial_l\sigma+\partial_k\sigma \partial_l\gamma + \partial_k\gamma \partial_l\gamma \right)\right\} \, .
\end{equation}

Considering now that both $\gamma$ and $\sigma$ are functions of a scalar $\varphi(x^k)$ (that is independent on the cyclic coordinate), one obtains 
\begin{eqnarray}\label{Ricci with psi}
    \mathcal{R} = e^{\gamma + \sigma} \sqrt{-|{g}_{kl}|}\Big\{ R[{g}_{kl}] + 2g^{kl}\Big[ (2\sigma' + \gamma') \nabla_l \nabla_k\varphi 
	\nonumber\\    
    + \left(2\sigma'' + \gamma'' + \sigma'^2 + \gamma'^2 +\sigma'\gamma' \right)\partial_k\varphi \partial_l\varphi \Big]\Big\} \, ,
\end{eqnarray}
where the prime represents a derivative with respect to $\varphi$. With the variational principle being the primary tool for our deduction, it is useful to express this scalar curvature density in terms of an ordinary divergence,  which does not contribute to the field equations. Following the same reasoning, it can be asserted that discrepancies originating from the parameter $\mathcal{R}$ may be disregarded, given their non-contribution to the field equations whenever $\mathcal{R}$ assumes an additive role within the Lagrangian framework. 

The reduction is made possible by the exponential $e^{\sigma(\varphi)+\gamma(\varphi)}$ in the following way:
\begin{equation}
    \nabla_l \nabla_k e^{\sigma(\varphi)+\gamma(\varphi)} = \left\{\left(\sigma'+\gamma' \right)\nabla_l \nabla_k \varphi + \left(\sigma''+\gamma'' + (\sigma' + \gamma' )^2\right) \partial_k \varphi \partial_l \varphi \right\} e^{\sigma(\varphi)+\gamma(\varphi)} \, ,
\end{equation}
which can be written as
\begin{equation}
   \left(\nabla_l \nabla_k \varphi \right) \ e^{\sigma(\varphi)+\gamma(\varphi)} = \frac{\nabla_l \nabla_k e^{\sigma(\varphi)+\gamma(\varphi)}}{(\sigma'+\gamma' )} 
    - \left[\frac{\sigma''+\gamma''}{\sigma'+\gamma'} + (\sigma'+\gamma') \right]\partial_k \varphi \partial_l \varphi \, e^{\sigma(\varphi)+\gamma(\varphi)} \, .
\end{equation}
The factor in square brackets on the right-hand side, when multiplied by the metric $g_{kl}$, is only an ordinary divergence if the $\sigma'+ \gamma'$ term is a constant. This means that $\sigma'$ and $\gamma'$ must be constant, which implies that $\sigma$ and $\gamma$ are linear in $\varphi$, so $\sigma'' = 0$ and $\gamma'' = 0$, so that we have
\begin{equation}
   \left(\nabla_l \nabla_k \varphi \right) e^{\sigma(\varphi)+\gamma(\varphi)}= \nabla_l \nabla_k \left(\frac{ e^{\sigma(\varphi)+\gamma(\varphi)}}{(\sigma'+\gamma' )}\right) - (\sigma'+\gamma' )\partial_k \varphi \partial_l \varphi \ e^{\sigma(\varphi)+\gamma(\varphi)}\, .
\end{equation}

Introducing this expression in Equation \eqref{Ricci with psi} leads to 
\begin{eqnarray}
    \mathcal{R} &= & e^{\gamma + \sigma} \sqrt{-|{g}_{kl}|}\Big\{ R[{g}_{kl}] - 2g^{kl}\big(\sigma'^2+2\sigma' \gamma')\partial_k\varphi \partial_l\varphi \Big\} 
	\nonumber \\    
    && + g^{kl} \nabla_l \nabla_k \left[\left(2 + \frac{(2\sigma')}{(\sigma'+\gamma')}\right) e^{\sigma(\varphi)+\gamma(\varphi)}\right] 
    \nonumber \\
    &\doteq & e^{\gamma + \sigma} \sqrt{-|{g}_{kl}|}\Big\{ R[{g}_{kl}] - 2g^{kl}\big(\sigma'^2 +2\sigma' \gamma')\partial_k\varphi \partial_l\varphi \Big\} \ ,
\end{eqnarray}
where $\doteq$ represents an equality between two expressions that only differ from an ordinary divergence. Without the possibility to remove such a divergence, it is not possible to construct this method because the curvature density does not end up in a way that allows one to build a relation with the Lagrangian in Equation \eqref{Lagrangian Lambda}.

Considering that $\sigma = \alpha \varphi$ and $\gamma = (1-\alpha) \varphi$, the previous equation can be reformulated as
\begin{equation}\label{Ricci(alpha)}
    \mathcal{R}(\alpha) \doteq e^{\varphi} \sqrt{-|{g}_{kl}|}\Big\{ R[{g}_{kl}] - 2g^{kl}\big(\alpha(2-\alpha)\big)\partial_k\varphi \partial_l\varphi \big)\Big\} \,,
\end{equation}
which for $\alpha = 0$ reduces to
\begin{equation}
     \mathcal{R}(0) \doteq e^{\varphi} \sqrt{-|{g}_{kl}|} \{R[{g}_{kl}]\}\, , 
\end{equation}
allowing one to rewrite Equation \eqref{Ricci(alpha)} as
\begin{equation}
    \mathcal{R}(\alpha) + 2g^{kl}\big(\alpha(2-\alpha)\big)\partial_k\varphi \partial_l\varphi  \doteq \mathcal{R}(0) \, ,
\end{equation}
so that
\begin{equation}\label{Varional Principle}
    \int d^4x \, \biggr[  \mathcal{R}(\alpha) + 2e^\varphi g^{kl}\sqrt{-|g_{kl}|}\big(\alpha(2-\alpha)\big)\partial_k\varphi \partial_l\varphi \big) \biggr] \equiv  \int d^4x \, \mathcal{R}(0) \, ,
\end{equation}
for all variations of $g_{kl}$ and $\varphi$ that vanish at the boundary.


In Buchdahl's work, it is imposed that   $(-e^{2\varphi},g_{kl})$ ($g_{ab}$ metric, from Equation \eqref{metric static g_ab}, with $\alpha = 0$) is a static solution of the field equations for empty space. This renders the right side of the previous equation equal to zero, allowing one to derive the solutions for the field equations obtained from the Lagrangian 
\begin{equation}\label{Langrangian vacuum}
    \mathcal{L} =  R - \mu g^{ab} \partial_a\psi \partial_b \psi \, ,
\end{equation}
leading to the result from \cite{Buchdahl:1959nk}.

\subsection{Buchdahl's Method with $\Lambda$}

To obtain the solutions for the field equations derived from the Lagrangian given in Equation \eqref{Lagrangian Lambda} using Buchdahl's method, it is essential to introduce two key steps in the derivation process. The first step consists in subtracting $\int d^4 x \ 2\Lambda \sqrt{-|g_{ab}|}$ from both sides of Equation \eqref{Varional Principle}, resulting in 


\begin{eqnarray}\label{Varional Principle Lambda}
     \int d^4x \, \left[  \mathcal{R}(\alpha) + 2e^\varphi g^{kl}\sqrt{-|g_{kl}|}\big(\alpha(2-\alpha)\big)\partial_k\varphi \partial_l\varphi \big)    
     -2\Lambda \sqrt{-|g_{ab}|} \, \right]
		\nonumber\\
     = \int d^4x \, \biggr[ \mathcal{R}(0) -2\Lambda \sqrt{-|g_{ab}|} \biggr] \, ,
\end{eqnarray}
for all variations of $g_{kl}$ and $\varphi$ that vanish at the boundary. As in the previous case, it is necessary to render the right side of the equation equal to zero. However, by introducing the new term, the second step needed to obtain the solutions desired is to impose the condition that the $(g_{kl}, e^{2\varphi})$ metric must be a static solution of the field equations
\begin{equation}\label{Vaccum with lambda}
    R_{ab} = \Lambda g_{ab} \, ,
\end{equation}
 instead of the 
 vacuum solutions, therefore making the right-hand side of Equation \eqref{Varional Principle Lambda} equal to zero. Thus, Equation \eqref{Varional Principle Lambda} becomes 
\begin{equation}\label{Varional Principle Lambda2}
     \int d^4x \, \left[  \mathcal{R}(\alpha) + 2g^{kl}e^{\varphi}\sqrt{-|g_{kl}|}\big(\alpha(2-\alpha)\big)\partial_k\varphi \partial_l\varphi \big) -2\Lambda \sqrt{-|g_{ab}|} \right] = 0 \, ,
\end{equation}
which, by using Equation \eqref{relação determinantes g}, can be written as
\begin{equation}\label{Varional Principle Lambda final}
     \int d^4x \, \sqrt{-|g_{ab}|} \left[ R(\alpha)-2\Lambda + e^{-2\alpha\varphi}g^{kl}\big(2\alpha(2-\alpha)\big)\partial_k\varphi \partial_l\varphi \big)  \right] = 0 \, .
\end{equation}

One must keep in mind that the field equations that arise from Equation \eqref{Lagrangian Lambda} will be satisfied by the functions $g_{ab}$ ($g_{ab} \equiv ('g_{kl},'g_{00})$) and $\psi$ if 
\begin{equation}\label{Varional Principle original}
     \int d^4x \, \sqrt{-|g_{ab}|}\left[ R -2\Lambda - \mu g^{ab}\partial_a\psi \partial_b\psi \right] = 0 \, ,
\end{equation}
for arbitrary variations that disappear at the boundary. By comparing Equation \eqref{Varional Principle Lambda final} with Equation  \eqref{Varional Principle original}, it is clear that by considering $'g_{kl} = e^{2\alpha\varphi}g_{kl}$, $'g_{00} = -e^{2(1-\alpha)\varphi}$, $\varphi = \frac{\psi}{2\kappa}$, where $\kappa$ is an arbitrary constant and $\alpha =(1 \pm \sqrt{1+2\mu \kappa^2}) $ by Equation \eqref{Varional Principle Lambda final}, the condition \eqref{Varional Principle original} is~satisfied.

With this result, one concludes that the field equations from the Lagrangian \eqref{Lagrangian Lambda} have a one-parameter set of pairs of static solutions given by
\begin{eqnarray}
    g_{ab} &=& \left(-(-g_{00})^{1-\alpha},(-g_{00})^{\alpha} g_{kl}\right)\label{metric solution} \, , \\ 
    \psi &=& \kappa \ln(-g_{00})\label{scalar solution} \, ,\\
    \alpha &=& 1 \pm \sqrt{1+2 \mu \kappa^2}  \label{constant solution}\,,
\end{eqnarray}
respectively.
  The metric $g_{ab}$ of Equation \eqref{Lagrangian Lambda} is a solution of the field equations of the action defined by this Lagrangian, and $g_{00}$ and $g_{kl}$ are the components of the metric, which is a solution of the field Equation \eqref{Vaccum with lambda}. These static solutions remain valid for the field equations derived from \eqref{Langrangian vacuum}, contingent upon the condition that the metric with the components $g_{00}$ and $g_{kl}$ satisfies the vacuum Einstein field equations.
%


\section{Applications}\label{Sec4:applications}

\subsection{Scalar-Tensor Gravity Theory}

As previously exposed, an appropriate conformal transformation of the space-time metric has the ability to map the ST theory onto GR, plus a scalar field that is coupled to matter.  If we consider a vacuum configuration, this means that the ST, and in particular the Brans--Dicke solution,  is applied into GR plus a single scalar field, which happens to be the redefinition of the  scalar field that originally couples to the curvature of the spacetime. We say that we have transformed the theory from the original Jordan frame to a target  Einstein~frame.

Thus, Buchdahl's result on reciprocal static metrics \cite{Buchdahl:1959nk} conveys 
a simple correspondence between vacuum solutions from two alternative gravity theories, namely the ST and GR theories. Indeed, as it establishes the correspondence between a vacuum GR solution and a reciprocal GR solution with a massless scalar field, the latter can be interpreted as the Einstein frame representation of a vacuum Brans--Dicke solution~\cite{O'Hanlon+Tupper 72} in the Jordan frame. If we have a solution of vacuum GR, we can use the latter device based on the conformal rescaling to generate a solution of the ST theory.

The generalized Brans--Dicke scalar-tensor field equations are obtained from the action~\cite{Will:2014kxa,Mimoso:1993dphil}
\begin{equation}\label{GBD}
    S = \frac{1}{16\pi} \int d^4x \sqrt{-g} \left[\phi R - \frac{\omega(\phi)}{\phi} \partial_a \phi \partial_b \phi + 16\pi \mathcal{L}_m \right] \, ,
\end{equation}
where \emph{R} is the Ricci curvature scalar of the spacetime associated with the $g_{ab}$ metric, $\phi$ is the Brans--Dicke scalar field, $\omega(\phi)$ is the coupling parameter, and $\mathcal{L}_m$ represents the Lagrangian for the matter sector. 

Varying the action \eqref{GBD} with respect to the two dynamical variables $g_{ab}$ and $\phi$ leads to
\begin{equation}\label{GBD variation metric}
    R_{ab} - \frac{1}{2}g_{ab}R = 8\pi \frac{T_{ab}}{\phi}+ \frac{1}{\phi} (\nabla_a\nabla_b \phi - g_{ab}\square\phi) + \frac{\omega(\phi)}{\phi^2} \left(g^c_a g^d_b -\frac{1}{2} g_{ab}g^{cd} \right) \partial_c\phi \partial_d\phi \, ,
\end{equation}
and
\begin{equation}\label{kG equation}
   \square\phi =\frac{1}{2\omega(\phi)+3} \left( 8\pi T - g^{cd} \partial_c \omega \partial_d \phi \right) \, ,
\end{equation}
respectively, 
where $T\equiv T^a_a$ is the trace of the energy–momentum tensor of the matter defined as
\begin{equation}
T^{ab} = \frac{2}{\sqrt{-g}}\frac{\partial}{\partial g_{ab}}(\sqrt{-g} \mathcal{L}_m) \, .
\end{equation}

It is known that a theory featuring a gravitational coupling that fluctuates, such as ST gravity, can be seen as being equivalent to another theory where the gravitational coupling remains constant, but masses and lengths exhibit variation \cite{Dicke:1961gz}. This equivalence can be demonstrated mathematically through the utilization of a conformally rescaled metric. By considering the following conformal transformation \cite{Mimoso:1994wn}
\begin{equation}\label{conformal transformation}
    \hat{g}_{ab} = \frac{\phi}{\phi_0} g_{ab} \, ,
\end{equation}
where $\phi_0$ is an arbitrary constant introduced to keep the conformal factor dimensionless, the action from Equation \eqref{GBD} becomes
\begin{equation}
    S = \frac{1}{16\pi} \int d^4x \sqrt{-\hat{g}}  \left[\frac{1}{\phi_0}\hat{R}  +16\pi \left( - \frac{1}{2} \partial_a \psi \partial_b \psi + \left(\frac{\phi_0}{\phi}\right)^2\mathcal{L}_m \right) \right] \, ,
\end{equation}
where we introduce a new scalar field $\psi(\phi)$ defined by
\begin{equation}
    d\psi = \sqrt{\phi_0\frac{ (3+2\omega(\phi))}{16\pi}} \frac{d\phi}{\phi} \, ,
\end{equation}
With this transformation and considering $\phi_0 \equiv G^{-1}$, the action \eqref{GBD} is reduced simply to the Einstein–Hilbert action of GR with a scalar field, being called the Einstein frame, while   Equation \eqref{GBD} is denoted by the Jordan frame. 

\subsection{Scalar-Tensor Gravity Plus a Massless Scalar Field}
In the Jordan frame, when the content of the matter Lagrangian is a massless scalar-field such as
\begin{equation}
\mathcal{L}_m = -\frac{1}{2} \partial_a \chi \partial_b \chi \, ,
\end{equation}
that has the wave equation
\begin{equation}
\Box \chi = 0 \ .
\end{equation}

The conformal transformation to the Einstein frame leads to a composite Lagrangian for the matter massless scalar field and the transformed scalar-field given by
\begin{eqnarray}
\mathcal{L}_{\phi + \chi} &=& -\frac{1}{2} \hat{g}_ {ab} \left(\frac{\partial_a \chi \partial_b \chi}{G\phi} + \partial_a \psi \partial_b \psi \right) 
	\nonumber \\
&=& -\frac{1}{2} \hat{g}_ {ab}\left(\frac{\partial_a \chi \partial_b \chi}{G\phi} + \frac{3+2w(\phi)}{16\pi G}\frac{\partial_a \phi \partial_b \phi}{\phi^2}\right) \, .
\end{eqnarray}

If the fields are a function of a single variable, which is a condition imposed by spherical symmetry or homogeneity, then $\chi$ can be written as a function of $\phi$, allowing one to define the composite scalar 
\begin{equation}\label{varphi definition}
	d\varphi = \sqrt{\frac{3+2\Tilde{\omega}(\phi)}{16\pi G}} \frac{d\phi}{\phi} \, ,
\end{equation} 
where 
\begin{equation}\label{omega tilde}
\Tilde{\omega}(\phi) = \omega(\phi) + 8\pi \phi \left(\frac{d \chi}{d \phi}\right)^2 \, .
\end{equation}

Therefore, the action of this ST theory in the Einstein frame is
\begin{equation}
S = \frac{1}{16\pi G} \int d^4 x \sqrt{-|\Hat{g}|} \left[\hat{R}+ 16\pi G \left(-\frac{1}{2} \hat{g}^{ab} \partial_a \varphi \partial_b \varphi \right) \right] \, . 
\end{equation}

It is important to emphasize that the Einstein metric should constitute a solution for the massless field/stiff fluid. In other words, solving the field equations for the massless scalar, we have~\cite{Wands,Yazadjiev:2001bx}
\begin{equation}
\Hat{\Box} \chi = \Hat{g}^{ab} \frac{\partial_a\phi \partial_b\chi}{\phi} \, .
\end{equation}

These results establish a relationship between a ST theory with matter in the Jordan frame and a corresponding vacuum conformal scalar-tensor theory in the Einstein frame. This conclusion aligns with the demonstration in \cite{Mimoso:1994wn}, which showed that, within a Friedmann-Lemaître-Robertson-Walker (FLRW) cosmological description, assuming a perfect fluid described by the barotropic equation of state $ p = (\gamma - 1)\rho $, the specific case where $\gamma = 2$, corresponding to stiff matter (characterized by the long wavelength modes of a massless scalar field), exhibits an equivalence to the vacuum case.

Thus, it is possible to map a solution from the Einstein frame to the Jordan frame, even in the presence of matter in the latter. (An alternative derivation of scalar-tensor solutions with a massless field was done in \cite{Yazadjiev:2001bx}.) Applying the result of Equation \eqref{scalar solution} within the framework of a spherically symmetric metric, which satisfies the vacuum field equations of GR, inherently ensures that the scalar fields are functions of a single variable. Consequently, this allows for the derivation of an expression for the composite scalar field that is solely dependent on the metric used, given by 
\begin{equation}\label{varphi equation}
	\varphi = \kappa \ln(-g_{00}) \, .
\end{equation}

In the scenario where $\Tilde{\omega}$ is constant, the composite scalar $\chi$ is given by the differential~equation
\begin{equation}\label{Differential equation chi}
    \frac{d\chi}{d\theta} = \pm \sqrt{\frac{\Tilde{\omega} -\omega(\phi)}{8\pi \phi}} \, ,
\end{equation}
and, in conjunction with Equations \eqref{varphi definition} and \eqref{varphi equation}, leads to 
\begin{equation}\label{phi massless}
	\phi= (-g_{00})^{\kappa\sqrt{16\pi G/ 3+2\Tilde{\omega}}} \, .
\end{equation}

Furthermore, if the coupling parameter $\omega(\phi)$ of the Jordan frame is constant, denoted as $\omega_0$, then Equation \eqref{Differential equation chi} simplifies to

\begin{equation}
    \chi = \pm 2\eta \phi^{\frac{1}{2}} \, ,
\end{equation}
where $\eta$ is 

\begin{equation}
   \eta \equiv  \sqrt{\frac{\Tilde{\omega} - \omega_0}{8\pi}} \, .
\end{equation}

In addition, as shown in \cite{Wands}, the relation of the massless scalar field and the original scalar-field encodes two interacting stiff fluids that have a dynamical effect of a single perfect stiff fluid; in other words,
\begin{equation}
\rho = \rho_{\chi} + \rho_{\phi} \, ,
\end{equation}
with parallel velocities, $u=u_\chi=u_\phi$ (condition satisfied if $\chi$ is a function of $\phi$), and
\begin{align}
	p_{\chi} &= \rho_{\chi} = \frac{1}{2} \left|\hat{g}^{ab} \frac{\partial_a \chi \partial_b \chi}{G \phi}\right|\, , \\ 
	p_\phi &= \rho_{\phi} = \frac{1}{2} \left|\hat{g}^{ab} \frac{3+2w(\phi)}{16\pi G}\frac{\partial_a \phi \partial_b \phi}{\phi^2} \right| \, .
\end{align}

With the previous results for $\phi$ and $\chi$ for the case of $\Tilde{\omega}$ and $\omega$ constant, the perfect stiff fluid is an energy density given by
\begin{equation} 
	\rho = \frac{1}{2} \left( \left|\frac{\eta^2}{G} \frac{\partial_a\phi \partial_b\phi}{\phi^2}\right| +  \left|\frac{3+2\omega}{16\pi G} \frac{\partial_a\phi \partial_b\phi}{\phi^2}\right| \right) \, ,
\end{equation}
where $\phi$ is given by Equation \eqref{phi massless}.

\subsection{Brans--Dicke Theory with a Quadratic Potential}

The scalar-tensor field equations for the Brans--Dicke theory with a potential are obtained from the action
\begin{equation}\label{BD with V JF}
    S = \frac{1}{16\pi} \int d^4x \sqrt{-g} \left[\phi R  - \frac{\omega_{BD}}{\phi} \partial_a \phi \partial_b \phi - 2V(\phi) \right]  + S_M \, ,
\end{equation}
where \emph{R} is the Ricci curvature scalar of the spacetime associated with the $g_{ab}$ metric, $\phi$ is the Brans--Dicke scalar field, $\omega_{BD}$ is the dimensionless Brans--Dicke coupling constant, $V(\phi)$ is a scalar field potential, and $S_M$ is the action for the matter sector. 

Varying the action \eqref{BD with V JF} with respect to the two dynamical variables $g_{ab}$ and $\phi$ leads to
\begin{adjustwidth}{-\extralength}{0cm}
\begin{equation}\label{BD variation metric}
    R_{ab} - \frac{1}{2}g_{ab}R = 8\pi\frac{T_{ab}}{\phi}+ \frac{1}{\phi} \left(\nabla_a\nabla_b \phi - g_{ab}\square\phi\right) 
    + \frac{\omega_{BD}}{\phi^2} \left(\partial_a\phi \partial_b\phi -\frac{1}{2} g_{ab} \partial^c\phi \partial_c \phi \right) - \frac{g_{ab}}{\phi} V(\phi) \, ,
\end{equation}
\end{adjustwidth}
and
\begin{equation}\label{kG equation quadratic potential}
    2\omega_{BD} \frac{\square\phi}{\phi} - \frac{\omega_{BD}}{\phi} g^{ab} \partial_a\phi \partial_b \phi + R = 2\frac{\partial V}{\partial \phi} \, ,
\end{equation}
respectively.

Using the conformal transformation given by Equation \eqref{conformal transformation} and considering a vacuum, the action \eqref{BD with V JF} takes the form
\begin{equation}
    S = \frac{\phi_0}{16\pi} \int d^4x \sqrt{-\hat{g}} \left\{\hat{R}  - \frac{8\pi}{\phi_0} \partial_a \psi \partial_b \psi - U(\psi) \right\} \, ,
\end{equation}
where we introduce a new scalar field $\psi(\phi)$ defined by
\begin{equation}
    d\psi = \sqrt{\phi_0\frac{ (3+2\omega_{BD})}{16\pi}} \frac{d\phi}{\phi} \, ,
\end{equation}
and where 
\begin{equation}
    U(\psi) = \frac{2V \phi^3_0}{\phi^2} \, . 
\end{equation}

Once again considering $\phi_0 = G^{-1}$, the action \eqref{BD with V JF} is reduced to the Einstein–Hilbert action of GR plus a scalar field.

Considering $V(\phi) = \Lambda\frac{\phi^2}{\phi_0^3}$ leads to 
\begin{equation}\label{EF with V}
    S = \frac{1}{16\pi G} \int d^4x \sqrt{-\hat{g}} \left\{\hat{R} - 2\Lambda - 8\pi G \, \partial_a \psi \partial_b \psi  \right\} \, ,
\end{equation}
which is the action \eqref{Lagrangian Lambda} with $\mu=8\pi G$ (a different definition of the scalar $\psi$, due to different units for example, leads to a different value for the $\mu$ constant) that has a set of static solutions \eqref{metric solution} and \eqref{scalar solution} for the field equations that arise from varying the action with respect to $\hat{g}_{ab}$ and $\psi$.

Therefore, the static solutions for the field equations obtained from Equation \eqref{BD with V JF} with $V(\phi) = \Lambda\frac{\phi^2}{\phi_0^3}$ are related with the ones from Equation \eqref{EF with V} by the conformal factor leading~to
\begin{eqnarray}
    g_{ab} &=&  \left(-\phi_0(-\bar{g}_{00})^{1-\alpha-\frac{\kappa}{\beta}},\phi_0(-\bar{g}_{00})^{\alpha-\frac{\kappa}{\beta}}\bar{g}_{kl} \right)
    	\label{metric solution lambda} \, , \\ 
    \phi &=&  (-\bar{g}_{00})^\frac{\kappa}{\beta}
    	\label{scalar solution lambda} \, ,\\
    \alpha &=&\big(1 \pm \sqrt{1+16\pi G \kappa^2} \big) 
    	\label{constant solution lambda}\, ,
\end{eqnarray}
where   $g_{ab} = (g_{00},g_{kl})$ is the static metric that is the solution of the vacuum field equations derived from Equation \eqref{BD with V JF} with $V(\phi) = \Lambda\frac{\phi^2}{\phi_0^3}$,   $\bar{g}_{00}$ and $\bar{g}_{kl}$ are the components of the static metric $\bar{g}_{ab} = (\bar{g}_{00},\bar{g}_{kl})$, which must be a solution of the field Equation \eqref{Vaccum with lambda}, and\linebreak $\beta = \sqrt{\phi_0\frac{ (3+2\omega_{BD})}{16\pi}}$, which depends on the value of $\omega_{BD}$.

\subsubsection{The Schwarzschild--de Sitter Metric}

A case of particular interest is the known Schwarzschild--de Sitter metric \cite{Rindler:2006km}, described by the line element
\begin{equation}
    \textrm{d}s^2 = -\left(1-\frac{2m}{r} - \frac{\Lambda}{3}r^2\right) \textrm{d}t^2 
    + \left(1-\frac{2m}{r} - \frac{\Lambda}{3}r^2\right)^{-1}\textrm{d}r^2 + r^2 \textrm{d}\Omega^2 \, .
\end{equation}

Thus, by Equations \eqref{metric solution lambda} and \eqref{scalar solution lambda}, a solution of the field Equations \eqref{BD variation metric} and \eqref{kG equation quadratic potential}, with $V(\phi) = \Lambda\frac{\phi^2}{\phi_0^3}$ (with $\phi_0=1$), is given by
\begin{eqnarray}
    \textrm{d}s^2 = -\left(1-\frac{2m}{r} - \frac{\Lambda}{3}r^2\right)^{1-\alpha-\frac{\kappa}{\beta}} \textrm{d}t^2 
    + \left(1-\frac{2m}{r} - \frac{\Lambda}{3}r^2\right)^{1+\alpha-\frac{\kappa}{\beta}}\textrm{d}r^2 
    	\nonumber \\
    + \left(1-\frac{2m}{r} - \frac{\Lambda}{3}r^2\right)^{\alpha-\frac{\kappa}{\beta}}r^2 \textrm{d}\Omega^2 
    \, ,
\end{eqnarray}
and 
\begin{equation}
    \phi= \left(1-\frac{2m}{r} - \frac{\Lambda}{3}r^2\right)^{\frac{\kappa}{\beta}} \, .
\end{equation}

\subsubsection{The Nariai Solution}

Another special case is the neutral Nariai solution given by \cite{1950SRToh..34..160N}
\begin{equation}
    \textrm{d}s^2 = \Lambda^{-1} \left(-\sin^2(\chi) \textrm{d}\tau^2 + \textrm{d}\theta^2 + \textrm{d}\chi^2 +\sin^2(\theta)\textrm{d}\psi^2\right)\, ,
\end{equation}
where $\theta$ and $\chi$ both run from 0 to $\pi$, and $\psi$ has period 2$\pi$. For this metric, by choosing $\tau$ as the cyclic coordinate, the set of solutions is
\begin{equation}
    \textrm{d}s^2 = -\Big(\frac{\sin(\chi)}{\Lambda}\Big)^{1-\alpha-\frac{\kappa}{\beta}} \textrm{d}\tau^2 + \Lambda^{-1-\alpha+\frac{\kappa}{\beta}}\Big(\sin(\chi)\Big)^{\alpha-\frac{\kappa}{\beta}} (\textrm{d}\theta^2 + \textrm{d}\chi^2 +\sin^2(\theta)\textrm{d}\psi^2)\, ,
\end{equation}
and with 
\begin{equation}
    \phi = \Big(\frac{\sin(\chi)}{\Lambda}\Big)^{\frac{\kappa}{\beta}}\, .
\end{equation}

\subsubsection{Hyperbolically Foliated Solution}

It is also possible to apply the present extension of Buchdahl's result to hyperbolically foliated spacetimes.
For instance, consider the metric given by
\begin{equation}
ds^2= -e^{\mu(r)}\,{\rm d}t^2+ e^{\lambda(r)} \,{\rm
d}r^2+r^2\,({\rm d}u^2+\sinh^2{u}\,{\rm d}v^2) \,,
\label{metric_constnc}
\end{equation}
where the usual 2$-d$ spheres are replaced by pseudo-spheres,
${\rm d}\sigma^2={\rm d}u^2+\sinh^2{u}\,{\rm d}v^2$, that are
surfaces of negative, constant curvature \cite{Mimoso:2011eh,Lobo:2009du,Mimoso:2010yp,Stephani:2003tm}

In the presence of the cosmological constant $\Lambda$,  the general relativistic solution is
\begin{equation}
e^{\mu(r)}=e^{-\lambda(r)} =\left(\frac{\Lambda}{3}\,r^2+ \frac{2\mu}{r}-1\right) \; ,
\label{metric_constnc-antiScwarz}
\end{equation}
where $\mu$ is an integration constant.
Notice that this  solution  extends the one referred to as degenerate solutions of class A
\cite{Ehlers & Kundt 1962,Mimoso:2011eh,Stephani:2003tm}.


The scalar-tensor generalization of  the metric
(\ref{metric_constnc}) then becomes 
\begin{eqnarray}
{\rm d} s^2 &=& -\left(\frac{\Lambda}{3}\,r^2+\frac{2\mu}{r}-1\right)^B\,{\rm d}t^2+
\left(\frac{\Lambda}{3}\,r^2+\frac{2\mu}{r}-1\right)^{-B}\;{\rm d}r^2  \\ \nonumber
 & & \qquad  +\left(\frac{\Lambda}{3}\,r^2+\frac{2\mu}{r}-1\right)^{1-B}\,r^2\,({\rm d}u^2
+\sinh^2{u}\,{\rm d}v^2) \, , \label{ST_sol_constnc} \\
\varphi(r) &=& \sqrt{\frac{C^2(2\omega+3)}{16\pi}\varphi_0}\,
\ln\left(\frac{2\mu}{r}-1\right)  \; , \label{ST_sol_constnc_phi}
\end{eqnarray}
where
\begin{equation}
C^2= \frac{1-B^2}{2\omega+3} \qquad -1 \le B \le 1 \;.
\end{equation}
This clearly reduces to 
the GR limit when $B=1$, and hence $C=0$,
implying that $G=\Phi^{-1}$ is constant. 
This solution like the previous ones also exhibits two branches corresponding to $C =\pm
\{(1-B^2)/(2\omega+3)\}^{1/2}$.

As noted in \cite{Agnese:1985xj}, the $r=2\mu$ limit represents a true singularity, rather than just a coordinate singularity, which is evident from the analysis of the curvature invariants. In the spherically symmetric case, it was shown in \cite{Agnese:1985xj} that the singularity at $r=2\mu$ manifests as a point, causing the event horizon of the black hole to shrink to a point. In the Einstein frame, this occurs because the energy density of the scalar field diverges. In the scenario under consideration, the $r=2\mu$ condition now corresponds to the areal radius of the pseudo-spheres, $R=\left(2\mu/r-1\right)^{(1-B)/2},r$, which become zero.

Reverting $\varphi=\int\sqrt{\Phi_0(2\omega+3)/(16\pi)}\,{\rm
d}\ln(\Phi/\Phi_0)$, and the conformal transformation,
$g_{ab}=(\frac{\Lambda}{3}\,r^2 +2\mu/r-1)^{-C}\,\Tilde{g}_{ab}$, we can recast this
solution in the original frame in which the scalar-field is
coupled to the geometry and the content is a vacuum, i.e., the Jordan
frame. We derive the following solution
\begin{eqnarray}
\Phi(r) = \Phi_0\,
\left(\frac{\Lambda}{3}\,r^2+ \frac{2\mu}{r}-1\right)^C  \; , \label{ST_sol_JF_constnc_phi} 
\end{eqnarray}
and
\begin{eqnarray}
{\rm d} s^2 &=& -\left(\frac{\Lambda}{3}\,r^2+\frac{2\mu}{r}-1\right)^{B-C}\,{\rm d}t^2+
\left(\frac{\Lambda}{3}\,r^2+\frac{2\mu}{r}-1\right)^{-B-C}\;{\rm d}r^2 \nonumber\\
 & & \qquad +\left(\frac{\Lambda}{3}\,r^2+\frac{2\mu}{r}-1\right)^{1-B-C}\,r^2\,({\rm d}u^2
+\sinh^2{u}\,{\rm d}v^2) \, . \label{ST_sol_JF_constnc}
\end{eqnarray}
The  gravitational constant $G =\Phi^{-1}$  decays
from an infinite value at $r=0$ to  a vanishing value at $r=2\mu$
when $C>0$, and conversely, grows from zero at $r=0$ to become
infinite at $r=2\mu$, when $C>0$.

\section{Summary and Discussion}

The Brans--Dicke theory, introduced in 1961, modifies GR by incorporating a scalar field $\phi$ alongside the tensor field $g_{\mu\nu}$, and can be extended to the Scalar-Tensor (ST) theory by allowing for more general coupling functions and potential terms. This extension provides a more flexible framework for describing gravitational interactions and has significant implications for cosmology. A notable application of ST theory is Buchdahl's work on reciprocal static metrics and scalar fields within GR. Buchdahl's studies from the 1950s explored metrics with symmetry under reciprocal transformations, providing explicit solutions to the gravitational field equations. His techniques were instrumental in constructing a two-parameter family of solutions, including the Schwarzschild solutions as a special case, revealing intricate relationships between scalar fields and spacetime~geometry. 

Buchdahl's contributions have inspired extensive research over the years, and in this work, we have extended Buchdahl's method to include the cosmological constant and applied it to scalar-tensor cases beyond the traditional Brans--Dicke theory. More specifically, through a conformal transformation of the metric, we obtained new solutions for a version of Brans--Dicke theory incorporating a quadratic potential. 
We have also relaxed the application of Buchdahl's reciprocity to solutions of scalar-tensor theories that are not constrained to a vacuum, and include a minimally coupled massless scalar field in the Jordan frame.  The latter is equivalent to a stiff matter component and hence may be seen as an extension of Buchdahl's  method of generating scalar-tensor solutions.

In future research, we will further explore these extensions and physical issues that emerge in scalar vacuum solutions~\cite{Bronnikov:2001ah}. In particular, we will investigate in greater detail the solutions, as well as case studies  that were here presented mainly as illustrations. A more specific and complete analysis of each class of solutions was beyond the scope of the present work, devoted to the method of generating solutions. We will also envisage the application of Buchdahl's program to other classes of modified gravity theories, and to metrics endowed with spacelike Killing vectors that are of cosmological interest~\cite{MacCallum:1973kva,Ellis:1968vb,MacCallum:1993zqj}.
Further investigations of the results devised by Buchdahl's program  seem likely to open prospects for providing deeper insights into the nature of gravity and the dynamics of the~universe.

\vspace{6pt}
\authorcontributions{Conceptualization, D.S.P., J.P.M. and F.S.N.L.; methodology, D.S.P., J.P.M. and F.S.N.L.; validation, D.S.P., J.P.M. and F.S.N.L.; formal analysis, D.S.P., J.P.M. and F.S.N.L.; investigation, D.S.P., J.P.M. and F.S.N.L.; writing---original draft preparation, D.S.P., J.P.M. and F.S.N.L.; writing---review and editing, D.S.P., J.P.M. and F.S.N.L.; supervision, J.P.M. and F.S.N.L.; project administration, J.P.M.; funding acquisition, J.P.M. and F.S.N.L. All authors have read and agreed to the published version of the manuscript.}

\funding{This research was funded by the Funda\c{c}\~{a}o para a Ci\^{e}ncia e a Tecnologia (FCT) from the research grants UIDB/04434/2020, UIDP/04434/2020, and PTDC/FIS-AST/0054/2021.}

\institutionalreview{Not applicable.}

\informedconsent{Not applicable.}

\dataavailability{Data are contained within the article.} 



\acknowledgments{F.S.N.L. also acknowledges support from the Funda\c{c}\~{a}o para a Ci\^{e}ncia e a Tecnologia (FCT) Scientific Employment Stimulus contract with reference CEECINST/00032/2018.}

\conflictsofinterest{The authors declare no conflicts of interest.} 

\begin{adjustwidth}{-\extralength}{0cm}

\reftitle{References}



%


\PublishersNote{}
\end{adjustwidth}
\end{document}